# Theoretical and experimental study of the correlation between pulsed light repetition frequency and electric field measurement

**KE DI,[1] CHENGLIN YE,[1] MEIHUI LIU[2], PENGFEI SHI[2], YU LIU[1], JUN HE[2*] AND JIAJIA DU[1*]**

*[1]Chongqing University of Post and Telecommunications, Chongqing 400065, China*
*[2]State Key Laboratory of Quantum Optics Technologies and Devices, Shanxi University, Taiyuan 030006, China*
**author dujj@cqupt.edu.cn*

**We propose a novel approach for improving the performance of Rydberg atom sensors by utilizing the repetition frequency of pulsed lasers, which has been validated through experimental testing. Rydberg atoms excited by pulsed lasers are influenced significantly by the repetition frequency of the pulsed laser on the Rydberg state population. As the number of Rydberg atoms increases, the measurement sensitivity of the sensor to external fields also increases, directly enhancing the performance of the sensor. This paper investigates the response of the sensor to the same electric field when the repetition frequency of the pulsed laser is at the MHz level, with a focus on its gain effects on the broadcast communication frequency bands of 66MHz and 88MHz. This research substantiates the distinctive benefits of pulsed light for Rydberg atom excitation, thereby enhancing the efficacy of the detection of feeble signals and introducing a new approach for the development of more sensitive atomic sensors.**

Rydberg atoms are a special type of atoms in which electrons are excited to high-energy states (Rydberg states), resulting in extremely large electron orbital radii and unique physical properties[1,2]. Rydberg atoms are highly desirable in the disciplines of quantum computing [3-5] and precision measurement [6-8] due to their robust interactions and substantial electric dipole moments. This is particularly true for the preparation and manipulation of quantum bits and microwave field measurement [9,10]. Nevertheless, their practical applications are still plagued by insurmountable defects, and preparation efficiency and selectivity have become critical factors that restrict their applications, despite their broad prospects [11-13].

Pulsed laser excitation is developing as a novel method for producing Rydberg atoms, offering enhanced preparation efficiency and superior selectivity compared to classic continuous-wave laser excitation [14-16]. Recently, researchers have used femtosecond laser pulses to demonstrate that molecular excitation to Rydberg states is primarily achieved through the multiphoton correlation mechanism between electrons and nuclei [17]. They have constructed a 480nm picosecond pulsed laser system for ultrafast excitation of rubidium atoms to Rydberg states, achieving a Rydberg excitation efficiency of approximately 90% [18]. Simultaneously, the dynamic process of electron recapture into Rydberg states has been experimentally confirmed for the first time under the influence of powerful ultrashort laser pulses [19].

Pulsed laser excitation of Rydberg atoms, as a novel scheme [20], offers significant advantages in precision measurements. It can complete excitation within an extremely short time [21,22], reducing interference caused by external field distortion and collisions, thus significantly improving measurement accuracy. The high repetition rate of pulsed lasers can enhance the signal-to-noise ratio, making the detection of weak signals more reliable [23]. Furthermore, the exact control of excitation wavelength and pulse duration when exciting Rydberg atoms with pulsed lasers facilitates the selective stimulation of particular Rydberg states, hence circumventing measurement complications associated with multi-state superposition. This selective excitation is essential for investigating the intricate interactions between atoms and external fields, hence enhancing sensitivity to exceedingly weak physical variables.

Using a 509nm pulsed light and an 852nm laser, we produced Rydberg atoms through two-photon resonant excitation. We varied the repetition frequency of the 509nm pulsed light to investigate the gain effect on the electric field measurement sensitivity based on Rydberg atoms, while maintaining the Rabi frequency of the 509nm pulsed light constant and applying the same radio-frequency electric field to Cs atoms. The universality of this effect was verified at the broadcast communication frequencies of 66MHz and 88MHz. The method we propose can effectively enhance the sensitivity of electric field measurements using Rydberg atoms, providing a new technical approach for precision measurements based on Rydberg atoms.

The energy level diagram of Rydberg states in Cesium atoms$\left({}^{133}Cs\right)$ is shown in Figure 1(a), consisting of the ground state $|g\rangle$, the intermediate state $|i\rangle$, and the Rydberg state $|r\rangle$. Two laser beams resonantly couple the ground state $|g\rangle$ and the Rydberg state $|r\rangle$. Let $\omega_p$ and $\omega_c$ denote the angular frequencies of the probe light and the coupling light; $\Delta_p$ and $\Delta_c$ represent the frequency detunings of the probe light and the coupling light relative to the atomic resonant transitions. The reason why the form of the Rabi frequency in this paper is different from the standard form is that the mathematical description of the classical electromagnetic field is different in this paper, and the physical meaning is the same. The corresponding Rabi frequencies are $\Omega_p = \mu_{21} E_p/(2\text{h})$ and $\Omega_c = \mu_{32} E_c/(2\text{h})$, where $\mu_{21}$ and $\mu_{32}$ are the electric dipole moments of the respective transitions, $E_p$ and $E_c$ are the peak electric field amplitudes of the probe and coupling fields, respectively [24,25].

The Hamiltonian for the interaction between atoms and light fields is:

$$H=H_a+H_l+H_{al} \tag{1}$$

Where $H_a$ is the atomic Hamiltonian, $H_l$ is the light field Hamiltonian, and $H_{al}$ is the interaction Hamiltonian.

Within the framework of the rotating-wave approximation and the dipole approximation, the Hamiltonian matrix for the three-level atomic system is expressed as:

$$H=\frac{\hbar}{2}\begin{pmatrix} 0 & \Omega_p & 0 \\ \Omega_p & -2\Delta_p & \Omega_c \\ 0 & \Omega_c & -2(\Delta_p+\Delta_c) \end{pmatrix} \tag{2}$$

The time progression of the density matrix can characterize the evolution of a three-level system, with its dynamics articulated by the Lindblad master equation:

$$\dot{\boldsymbol{\rho}}=-\frac{i}{\hbar}[H,\boldsymbol{\rho}]+\boldsymbol{L}(\boldsymbol{\rho}) \tag{3}$$

Where $\boldsymbol{L}(\boldsymbol{\rho})$ is the dephasing matrix, and $\boldsymbol{\rho}$ is the atomic density matrix:

$$\boldsymbol{\rho}=\begin{pmatrix} \rho_{11} & \rho_{12} & \rho_{13} \\ \rho_{21} & \rho_{22} & \rho_{23} \\ \rho_{31} & \rho_{32} & \rho_{33} \end{pmatrix} \tag{4}$$

In a three-level atomic system, the dephasing matrix $\boldsymbol{L}(\boldsymbol{\rho})$ for spontaneous emission is expressed as:

$$\boldsymbol{L}(\boldsymbol{\rho})=\begin{pmatrix} \gamma_{21}\rho_{22} & -\frac{\gamma_{21}}{2}\rho_{12} & -\frac{\gamma_{32}}{2}\rho_{13} \\ -\frac{\gamma_{21}}{2}\rho_{21} & \gamma_{32}\rho_{33}-\gamma_{21}\rho_{22} & -\frac{\gamma_{21}+\gamma_{32}}{2} \\ -\frac{\gamma_{32}}{2}\rho_{31} & -\frac{\gamma_{21}+\gamma_{32}}{2}\rho_{32} & -\gamma_{32}\rho_{33} \end{pmatrix} \tag{5}$$

In the formula, $\gamma_{32}$ is the decay rate from Rydberg state $|r\rangle$ to intermediate state $|i\rangle$, and $\gamma_{21}$ is the decay rate from intermediate state $|i\rangle$ to ground state $|g\rangle$.

To maintain the coherence and conciseness of the main text, the theoretical models related to the excitation dynamics of pulsed light involved in this paper have been compiled in the supplementary document.

As shown in Figure 1(b), we present the contributions of different repetition frequencies to the population of Rydberg state atoms. In the theoretical simulation, the Rabi frequency of the probe field is $\Omega_p/(2\pi)$ =10MHz, the Rabi frequency of the coupling field is $\Omega_c/(2\pi)$ =10MHz, the decay rate of the intermediate state is $2\pi\times$5.2MHz, the decay rate of the Rydberg state is $2\pi\times$0.8kHz.

In the frequency domain, pulsed light is equivalent to a coherent frequency comb, where the comb spacing is equal to the repetition frequency. This is the fundamental reason why it can excite atomic ensembles with different velocity groups, which also leads to an increase in the atomic population of the Rydberg state. To investigate the relationship between the repetition frequency of the pulsed light and the Rydberg-state atomic population, in our theoretical simulations, we kept the Rabi frequency of the pulsed light constant and varied the repetition frequency, observing how the Rydberg-state population changes with the repetition frequency of the pulsed light, as shown in Fig. 1(b). In the simulation results, the repetition frequency of the pulsed light is $f_R$ = 33,44,55,66,77,88,96 MHz, and the Rydberg-state atomic population increases with the repetition frequency of the pulsed laser until it tends to be relatively stable.

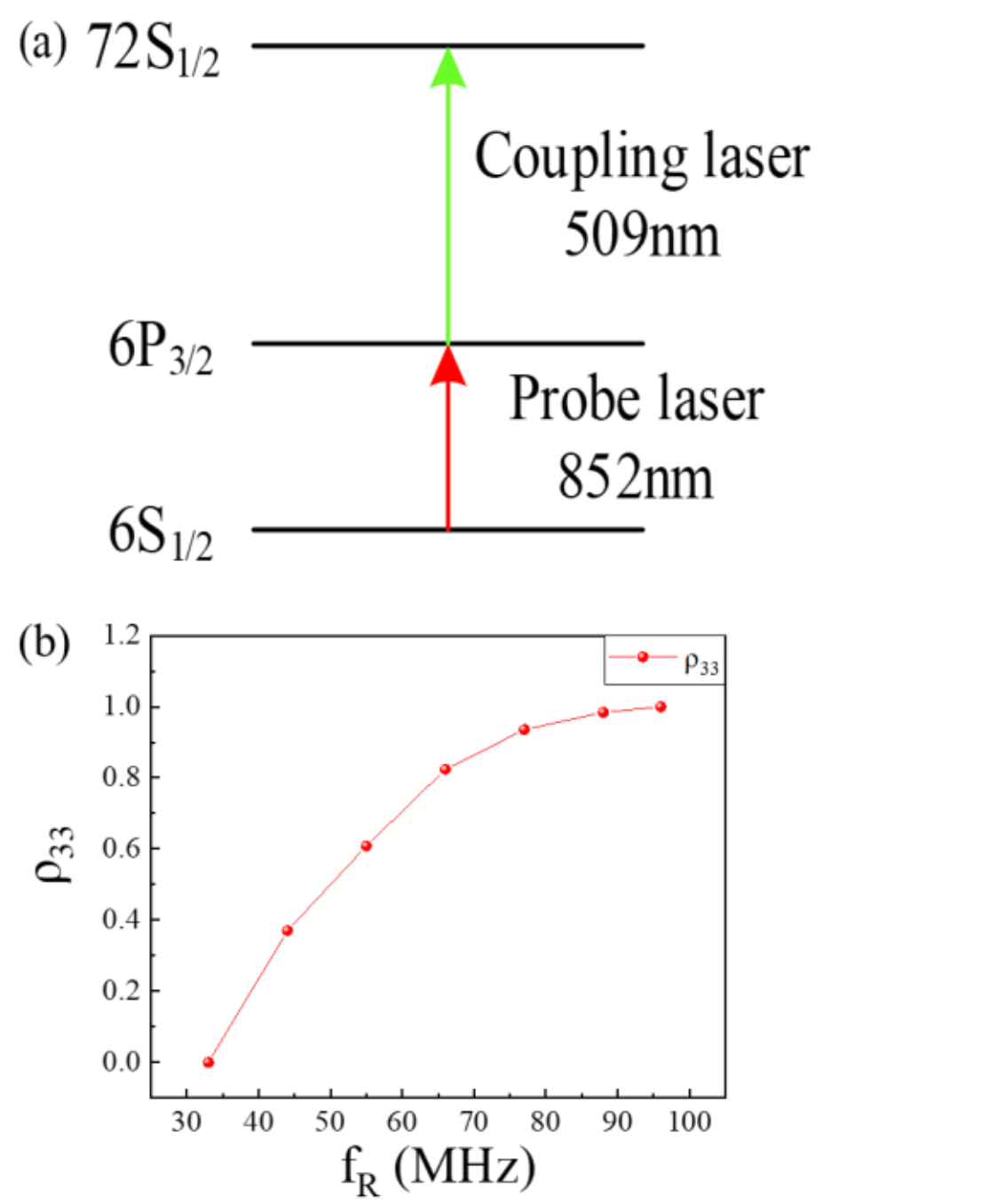

Fig. 1. (a) Energy level diagram of Rydberg EIT ladder for Cesium atoms. (b) The atomic population of the Rydberg state at pulse repetition frequencies $f_R$ =33,44,55,66,77,88,96 MHz.

After applying the signal field $E_{Sig}$ to the electrodes, the response of the Rydberg atoms to the electric field can be expressed as:

$$E_{tot}(t)=E_{Sig}e^{i(\omega_{Sig}t+\phi_{Sig})} \tag{6}$$

Where $E_{Sig}$ denotes the amplitude of the signal field, $\omega_{Sig}$ and $\phi_{Sig}$ represent the angular frequency and phase of the signal field, respectively. From the Stark shift formula, we can obtain:

$$\begin{aligned}\Delta f_{Stark}(t)&=-\frac{1}{2}\alpha[E_{tot}(t)]^2\\&=-\frac{1}{2}\alpha\left[E_{Sig}e^{i(\omega_{Sig}t+\phi_{Sig})}\right]^2\\&=-\frac{\alpha E_{Sig}^2}{4}-\frac{\alpha E_{Sig}^2}{4}\cos(2\omega_{Sig}t+2\phi_{Sig})\end{aligned} \tag{7}$$

Where $\alpha$ denotes the polarizability. The Rydberg electromagnetically induced transparency (EIT) signal oscillates at the frequency $E_{Sig}/\pi$, and its amplitude is linearly proportional to the signal field $E_{Sig}$. By measuring the amplitude variation of the oscillating signal, the measurement of the signal field can be achieved.

The experimental configuration is depicted in Fig. 2, including a radio-frequency (RF) electric field emission apparatus and an electric field measuring sensor utilizing Rydberg atoms. The RF electric field emission system consists of a function signal generator and two parallel electrode plates. The RF electric field signal generated by the function signal generator is guided to the two parallel electrode plates and finally applied to the Cs atoms. The core component of the sensor is an atomic vapor cell filled with cesium vapor. Utilizing two-photon resonance with a 509nm pulsed light and an 852nm laser, Cs atoms are effectively stimulated to Rydberg states, which are then employed for electric field sensing.

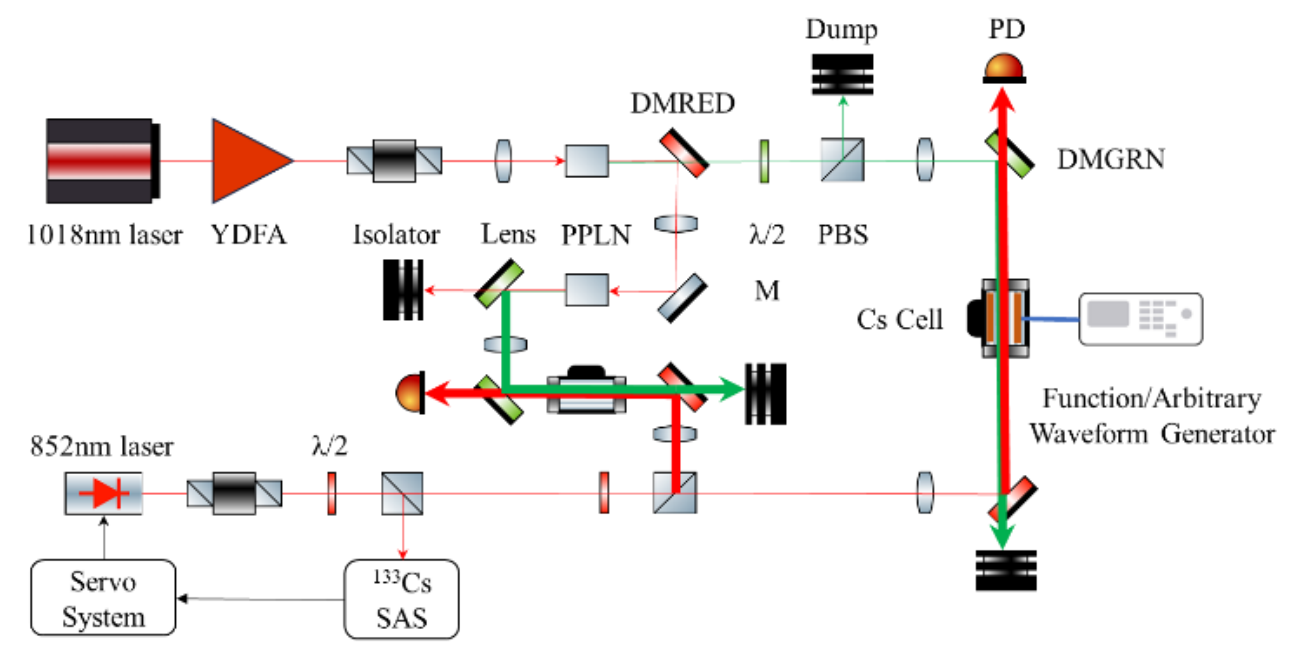

Fig. 2. Schematic of the experimental setup. YDFA: Ytterbium-doped fiber amplifier; Isolator: optical isolator; Lens: focusing lens; PPLN: periodically poled lithium niobate crystal (for frequency doubling); λ/2: half-wave plate; PBS: polarizing beam splitter; M: mirror; Cs Cell: cesium vapor cell; Dump: beam dump; PD: photodetector; DMRED / DMGRN: dichroic mirrors (reflecting red/green light, respectively); SAS: saturated absorption spectroscopy setup; Servo System: servo frequency stabilization system. The entire optical path is used to achieve Rydberg excitation and detection of $^{133}$Cs atoms.

In the experiment, we use an 852nm laser output by an external cavity diode laser (ECDL) as the probe light for real-time optical readout of the Rydberg states of cesium atoms, and lock its frequency at $6S_{1/2}(F=4) \rightarrow 6P_{3/2}(F=5)$. Conversely, the 509nm pulsed light employed as the coupling light in this experiment is frequency-doubled light that is produced by a self-constructed frequency doubling device. The internal pulse modulator of the laser converts the laser output from a 1018nm continuous laser seed source into pulsed light. This pulsed light is then amplified by a ytterbium-doped fiber amplifier (YDFA) and subsequently generated two beams of 509nm pulsed laser through a self-built two-stage frequency doubling system. Its frequency lies between the $6P_{3/2}$ (F=5) $\rightarrow 72S_{1/2}$ transition.

To finalize the measurement of the RF electric field, we drive a 509nm pulsed light beam and an 852nm laser to propagate in opposing directions within the Cs atomic vapor cell, therefore producing an EIT spectrum. This spectrum facilitates frequency stability of the 509 nm laser with a lock-in amplifier and a PID (Proportional-Integral-Derivative) controller. For the alternate beam, following the excitation of Cs atoms to Rydberg states, a radio frequency field is delivered via parallel electrode plates for the electric field measurement experiment.

To examine the influence of the pulsed light repetition frequency on electric field measurements utilizing Rydberg atoms, we sustained the pulse width of the pulsed light at 5 ns during the experiment. Simultaneously, to maintain a constant Rabi frequency of the 509nm pulsed light, it was necessary to modify the laser strength of the 509nm pulsed light in accordance with changes in the repetition frequency [26]. In the experiment, we ensured that the average power of the pulsed light increased proportionally to the repetition frequency to maintain the same single-pulse energy at different repetition frequencies. This verified that the laser intensity of the pulsed light remained constant per unit time throughout varying repetition frequencies, thereby maintaining the Rabi frequency of the pulsed light unaltered. In these settings, an RF electric field was applied to the cesium atomic vapor cell with constant frequency and amplitude, while the repetition frequency of the pulsed light was varied, and the signal-to-noise ratio (SNR) data of the measurement signals was recorded. Furthermore, the electric field measurement sensitivity can be reflected by measuring the variation of the SNR, as the measurement sensitivity is determined by dividing the field strength of the applied external field by the SNR at the corresponding measurement frequency. This is possible if the field strength of the external field remains constant during the experiment.

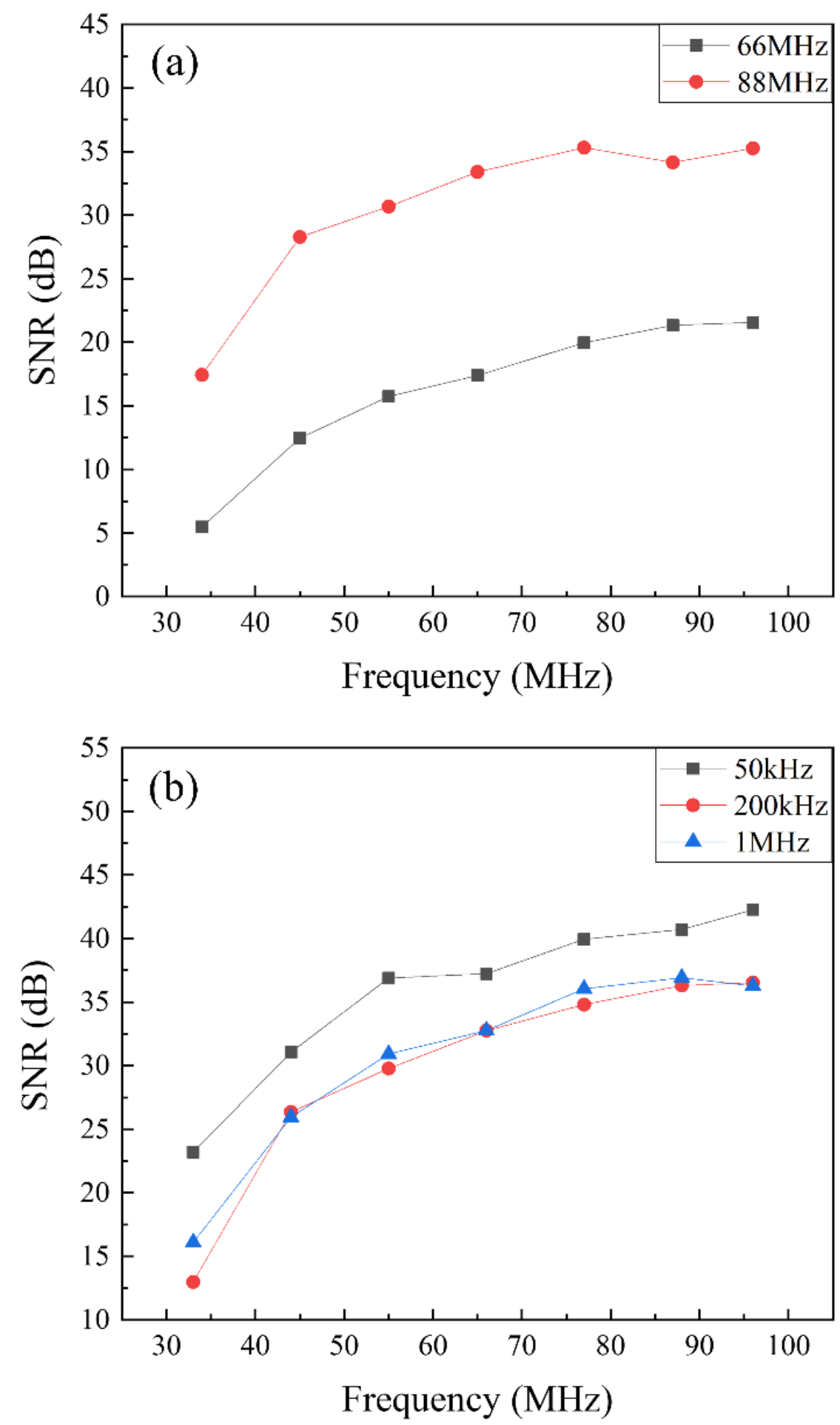

Fig. 3. (a) Variation trend of electric field measurement sensitivity with the repetition frequency of pulsed light when the RF fields are 50kHz, 200kHz, and 1MHz. (b) Variation trend of electric field measurement sensitivity with the repetition frequency of pulsed light when the RF fields are 66MHz and 88MHz.

As shown in Fig. 3(a), we measured the SNR of the measurement signals at RF electric field frequencies of 50 kHz, 200 kHz, and 1 MHz, respectively, under the same electric field amplitude. The experimental data show that the growth slope slows significantly beyond 80 MHz, indicating that the system enters a saturation regime. Meanwhile, the theoretical simulation in Fig. 1(b) clearly predicts that the population stabilizes after this frequency. Together, they support treating 80 MHz as the saturation onset for performance improvement, and this conclusion has been formulated with full consideration of the continuity in the data trend.

After theoretical analysis and comparison with the experimentally obtained data, we have reason to infer that when the pulse width is 5 ns and the repetition frequency exceeds 100

MHz, the electric field measurement sensitivity under the same conditions will also tend to be stable. The initial improvement originates from the increase in the number of effective excitations per unit time; the saturation phase marks that the system has reached a dynamic equilibrium between excitation and relaxation, and the maximum available Rydberg atomic population is limited, thereby determining the ultimate performance of the sensor device. It was noted that the SNR of the measurement signals increases by factors of 1.82, 2.81, and 2.25 as the repetition frequency of the pulsed light increases when the RF electric field frequencies are 50kHz, 200kHz, and 1MHz, respectively. Our initial conjecture and theoretical derivation results are substantially corroborated by this data. The experimental data's trend is in accordance with the theoretical simulation results, thereby confirming the accuracy of our theory and elucidating the pertinent impact of the pulsed light's repetition frequency on the electric field measurement using Rydberg atoms.

Building upon the aforementioned experiments, we further validated the universality of the theory. At the key broadcast communication frequencies of 66MHz and 88MHz, our experiments revealed that the theory remains applicable. As illustrated in the Fig. 3(b), using the same method at RF electric field frequencies of 66MHz and 88MHz, we measured the SNR of the measurement signals for Rydberg-atom-based RF electric field sensing across different pulsed light repetition frequencies. The results show that the SNRs increased by factors of 3.93 and 2.02, respectively, with increasing repetition frequency. This result verifies that at the key frequencies of broadcast communications, changes in the repetition frequency of pulsed light can significantly improve the Rydberg atom-based electric field measurement. Concurrently, our approach enhances the quantity of Rydberg state atoms, which can directly influence the signal strength, sensitivity, and spatial resolution of microwave electric field measurements. We believe this technology significantly enhances the measurement of the microwave electric field.

In summary, this paper investigates the impact of the repetition frequency of pulsed light on the population of Rydberg state atoms during the excitation of Rydberg atoms by pulsed light, and further explores the phenomenon that this influence enhances the performance of Rydberg atom sensors. In contrast to ns pulsed lasers, ps and fs pulsed lasers have the potential to further excite Rydberg atoms. Consequently, there is still space for further experimentation. The performance of precision measurement and sensing is substantially optimized [27] by increasing the number of sensing atoms in applications related to precision measurements based on Rydberg atoms, which is an effective approach to enhance measurement sensitivity [28,29]. Meanwhile, high-density Rydberg atoms exhibit advantages such as high sensitivity, wide frequency range, and strong anti-interference capability in broadcast communication, which contribute to improving communication quality and efficiency. Therefore, high-density Rydberg atoms demonstrate enormous potential in the fields of precision measurement and broadcast communication [30,31].

**Funding.** This work is supported by the Natural Science Foundation of Chongqing of China (Grant No.CSTB2024NSCO-MSX0746), State Key and the Laboratory of Quantum Optics and Quantum Optics Devices (Grant No.KF202408), the National Natural Science Foundation of China (Grant No. 52175531).

**Disclosures**. The authors declare no conflicts of interest.

**Data Availability Statement (DAS).** The data that support the findings of this study are available within the article.

**Supplemental Document.** See Supplement 1 for supporting content.